\documentclass[prc,preprint]{revtex4}
\usepackage{epsfig}
\usepackage{graphics}
\usepackage{amsmath}
\tighten

\renewcommand{\vec}[1]{\mathbf{#1}}

\begin{document}
\bibliographystyle{revtex}
\title{Separable RPA for self-consistent nuclear models}

\vspace{0.5cm}
\author{
V.O. Nesterenko$^1$, J. Kvasil$^2$ and P.-G. Reinhard$^3$
}
\affiliation{
$^{1}$ Bogoliubov Laboratory of Theoretical Physics,
Joint Institute for Nuclear Research
141980, Dubna, Moscow Region, Russia,
E-mail: nester@thsun1.jinr.ru\\
$^2$ Institute of Particle and Nuclear Physics, Charles
University, V.Hole\v sovi\v ck\'ach 2, CZ-18000 Praha 8, 
Czech Republic, E-mail: kvasil@ipnp.troja.mff.cuni.cz
\\
$^3$ Institut fur Theoretische Physik,Universitat Erlangen,
W-8520 Erlangen, Germany,
E-mail: mpt218@theorie2.physik.uni-erlangen.de
}
\date{\today}
\begin{abstract}
Self-consistent factorization of two-body residual interaction 
is proposed for arbitrary density- and current-dependent energy 
functionals. Following this procedure, a separable RPA (SRPA) 
method is constructed. SRPA considerably simplifies the 
calculations and demonstrates quick convergence to
exact results.  The method is tested for SkI3 and SkM* forces.
\end{abstract}
\maketitle

\section{Introduction}

Self-consistent nuclear models, as e.g. Skyrme-Hartree-Fock (SHF), 
are widely used for the description of nuclear ground state
properties. They do also allow a description of excitation
spectra. This is usually done within the random-phase approximation
(RPA), which is meanwhile a textbook standard in nuclear physics
\cite{Row,Ber94aB}, 
for a few recent applications see \cite{Rew,Col,Vre}. Actually,
these dynamical applications are up to now rather limited. The reason
is that RPA implies the diagonalization of large matrices whose rank
is determined by the size of the particle-hole $1ph$ space. The number
of $1ph$ states grows huge for heavy and/or deformed nuclei. This
limits the range of application for fully fledged RPA.
The most applications are thus found for spherical nuclei.

The RPA problem becomes much simpler if the residual two-body
interaction is reduced to a separable form, i.e.
\begin{eqnarray}
  \hat{V}_{\rm res}^{\mbox{}}  
  \rightarrow
  \hat{V}_{\rm res}^{\rm(sep)}
  &=&
 \frac{1}{2}
 \sum_{k,k'=1}^{N_{sep}} \mu_{kk'} {\hat Z}_k {\hat Z}_{k'} , 
 \nonumber
\\
 {\hat Z}_k
 &=&
  \sum_{ph} \langle p|{\hat Z}_k |h\rangle a^{\dagger}_p a_h\;
\label{eq:1}
\end{eqnarray}
where $\{{\hat Z}_k\}$ is a set of hermitian one-body operators.  The
factorization changes the rank of the RPA matrix from the number of
$1ph$ configurations to the number of basis operators ${\hat Z}_k$ and
this reduces dramatically the computational effort. The success of a
separable ansatz depends, of course, on a diligent choice of the
operators ${\hat Z}_k$ and the associated strength coefficients
$\mu_{kk'}$.

Factorization of the residual interaction is widely used in nuclear
theory but mainly within trivial schemes exploiting one separable term
with an intuitive guess for the separable one-body operator ${\hat
Z}$. The strength constant $\mu$ is usually fitted to reproduce
available experimental data (see e.g. \cite{Sol}). Obviously, accuracy
and predictive power of such simple schemes are limited.
Several improvements towards self-consistent schemes have been
proposed during the last decades \cite{Row}-\cite{Vor}.  However, these
schemes are not sufficiently general. Some of them are limited to
analytic or simple numerical estimates \cite{Row}-\cite{SS}, others
are not fully self-consistent in the sense that they start from
phenomenological single-particle potentials \cite{Ne_PRC}-\cite{Kubo}
or cover only particular effective forces \cite{Vor}.

In the present paper we propose a general self-consistent separable
RPA (SRPA) approach relevant to arbitrary density- and
current-dependent functionals.  The self-consistent scheme of
Ref. \cite{LS} is generalized to the case of several separable
operators. The operators are chosen to have maxima at different areas
of the nucleus. This is crucial for an accurate reproduction of the
residual interaction $V_{\rm res}$. A similar scheme has been 
successfully
applied to the Kohn-Sham functional for description of collective
oscillations of valence electrons in atomic clusters
\cite{Ne_PRA}-\cite{babst}. The nuclear case is more demanding since
the SHF functional is much more involved.  We will discuss the actual
SRPA scheme for the case of SHF and present first successful tests for
isoscalar E2 and isovector E1 giant resonances in $^{40}{\rm Ca}$ and
$^{208}{\rm Pb}$ using two typical Skyrme parameterizations.

\section{SRPA}

\subsection{Separable operators}

In connection with nuclear density functional theory, it is preferable
to sort the one-body operators ${\hat Z}_k$ according to time
parity. We thus decompose the set into hermitian time-even  ${\hat
X}_{k}$ and time-odd ${\hat Y}_{k}$ operators and the
corresponding strength matrices $\kappa_{kk'}$ and $\eta_{kk'}$.  
Then the separable Hamiltonian can be written as
\begin{eqnarray}
 \hat{H}_{\rm SRPA}
 &=&
 \hat{h}_0+\hat{V}_{\rm res}^{\rm(sep)} ,
\label{eq:hres}
\\
 \hat{V}_{\rm res}^{\rm(sep)}
 &=&
  \frac{1}{2}\sum_{kk'=1}^{N_{sep}}
 ( \kappa_{kk'}{\hat X}_{k}{\hat X}_{k'}
 +\eta_{kk'}{\hat Y}_{k}{\hat Y}_{k'})
\label{eq:vres}
\end{eqnarray}
where $\hat{h}_0$ is the mean field Hamiltonian of the ground state,
$\kappa_{kk'}$ and $\eta_{kk'}$ are symmetric matrices. Note
that the operator product is considered to be separable at the 
Hartree level, i.e. for an arbitrary $1ph$ operator $\hat{A}$
we have
\begin{equation}
  [\hat{Z}_k\hat{Z}_{k'},\hat{A}]
  =
  \hat{Z}_k\langle[\hat{Z}_{k'},\hat{A}]\rangle
  +
  \hat{Z}_{k'}\langle[\hat{Z}_{k},\hat{A}]\rangle .
\end{equation}
Furthermore, it is important to note for the following considerations
that only commutators between time-even and time-odd operators have
non-vanishing expectation values. That means that
$\langle[\hat{X}_k,\hat{X}_{k'}]\rangle=
 \langle[\hat{Y}_k,\hat{Y}_{k'}]\rangle=0$
while $\langle[\hat{X}_k,\hat{Y}_{k'}]\rangle\neq 0$.

The success of a separable ansatz depends very much on an appropriate
choice of the expansion basis ${\hat X}_{k}$ and ${\hat Y}_{k}$.
Collective modes (vibrational, giant resonances) dominate the spectral
strengths and it is intuitively clear that collective operators like
the various multipole moments should play the leading role in the
expansion. We will denote them as
hermitian time-even collective operators
$\hat{Q}_{k}$ (as e.g. a local multipole operator
$\hat{Q}_{\lambda\mu}({\vec r})\sim
r^{\lambda}(Y_{\lambda\mu}+Y^{\dagger}_{\lambda\mu})$) and
time-odd conjugate momenta
\begin{equation}
\hat{P}_{k}=i[\hat{H},\hat{Q}_{k}]_{ph}
\end{equation}
where the index $ph$ means the $1ph$ part of the operator.
These operators can serve as generators of a collective 
motion by virtue of the scaling transformation
\begin{equation}
  |\Psi\rangle
  =
  e^{\hat{G}}|\rangle ,
  \quad
  \hat{G}
  =
  \sum_k (ip_k\hat{Q}_k-iq_k\hat{P}_k)
\label{eq:scaling}
\end{equation}
where $|\rangle \equiv | \Psi_0\rangle$ is the ground state wave
function, $\hat{G}$ is the generator, composed of various
coordinate-like $\hat{Q}_k$ and momentum-like $\hat{P}_k$ operators,
$p_k$ and $q_k$ are the corresponding (time-dependent) amplitudes,
see e.g. \cite{Boh85,Bra89,ReiBraGen}. The dynamical deformation
(\ref{eq:scaling})
induces changes of the mean-field Hamiltonian $\hat{h}$ which can be
expanded in the linear regime for the small $\hat{G}$ as
\begin{equation}
  \begin{array}{rclcc}
   \hat{h}
   &\approx&
   \hat{h}_0
   &+&
   \underbrace{[\hat{h}_0,\hat{G}]_{ph} + 
   [\hat{V}_{\rm res},\hat{G}]_{ph}}_{}
  \\
  &=&
 \hat{h}_0 &+& [\hat{H},\hat{G}]_{ph} .
 \end{array}
\label{eq:hrespG}
\end{equation}\
The dynamics of small-amplitude vibrations is governed by
$[\hat{H},\hat{G}]_{ph}$ and so we optimize the separable 
interaction by requiring that
\begin{equation}
  [\hat{H},\hat{G}]_{ph}
  =
  [\hat{H}^{\rm(sep)},\hat{G}]_{ph}
\label{eq:HG}
\end{equation}
as close as possible. Since both Hamiltonians contain $\hat{h}_0$ in
the same way, the requirement (\ref{eq:HG}) is essentially
\begin{equation}
  [\hat{V}_{\rm res},\hat{G}]_{ph}
  =
  [\hat{V}_{\rm res}^{\rm(sep)},\hat{G}]_{ph} .
\label{eq:require}
\end{equation}
Applying (\ref{eq:require}) to generators
$\hat{Q}_{k}$ and $\hat{P}_{k}$ separately, we have
\newpage
\begin{subequations}
\begin{eqnarray}
  [\hat{V}_{\rm res},\hat{P}_k]_{ph}
   &=&
  [\hat{V}_{\rm res}^{\rm(sep)},\hat{P}_k]_{ph}
  \nonumber\\
  &=& \sum_{\tilde{k}k'}
  {\hat X}_{\tilde{k}}\kappa_{\tilde{k}k'}
  \langle[{\hat X}_{k'},\hat{P}_k]\rangle ,
\\{}
  [\hat{V}_{\rm res},\hat{Q}_k]_{ph}
  &=&
  [\hat{V}_{\rm res}^{\rm(sep)},\hat{Q}_k]_{ph}
\nonumber\\
  &=& \sum_{\tilde{k}k'}
  {\hat Y}_{\tilde{k}}\eta_{\tilde{k}k'}
  \langle[{\hat Y}_{k'},\hat{Q}_k]\rangle .
\end{eqnarray}
\end{subequations}
This shows that an optimal choice has to fulfill the
condition
\begin{subequations}
\begin{eqnarray}
  {\hat X}_k\in\left\{[\hat{V}_{\rm res},\hat{P}_k]_{ph}\right\}
,
\\
  {\hat Y}_k\in\left\{[\hat{V}_{\rm res},\hat{Q}_k]_{ph}\right\}
.
\end{eqnarray}
\end{subequations}
We are free to assume any linear recombination among the sets
$\{\hat{X}_k,\hat{Y}_k\}$ and thus we may choose
\begin{subequations}
\label{eq:sepform}
\begin{eqnarray}
\label{eq:Xgen}
  \hat{X}_{k}
  &=&
  [\hat{V}_{\rm res},\hat{P}_{k}]_{ph} ,
\\
\label{eq:Ygen}
  \hat{Y}_{k}%
  &=&
  [\hat{V}_{\rm res},\hat{Q}_{k}]_{ph} .
\end{eqnarray}
\end{subequations}
Substitution of $\hat{V}^{\rm(sep)}_{\rm res}$
into (\ref{eq:Xgen}) and (\ref{eq:Ygen})
has to give the identities. This results in the expressions for the
{\it inverse} strength constants $\kappa_{k'{\bar k}}^{-1}$ and
$\eta_{k'{\bar k}}^{-1}$
($\sum_{\bar k}\kappa_{k{\bar k}}^{-1}\kappa_{{\bar k}k'}=\delta_{kk'}$,
 $\sum_{\bar k}\eta_{k{\bar k}}^{-1}\eta_{{\bar k}k'}=\delta_{kk'}$):
\begin{subequations}
\label{eq:sepstreng}
\begin{eqnarray}
\label{eq:k_kk'}
  \kappa_{k'{\bar k}}^{-1}
  &=&
  \langle[{\hat X}_{k'},\hat{P}_{\bar k}]\rangle
  =
  \langle[[\hat{V}_{\rm res},\hat{P}_{k}]_{ph},\hat{P}_{\bar k}]\rangle
  ,
\\
\label{eq:e_kk'}
  \eta_{k'{\bar k}}^{-1}
  &=&
  \langle[{\hat Y}_{k'},\hat{Q}_{\bar k}]\rangle
  =
  \langle[[\hat{V}_{\rm res},\hat{Q}_{k}]_{ph},\hat{Q}_{\bar k}]\rangle
  .
\end{eqnarray}
\end{subequations}
The actual calculation of the commutators can be done either in
wavefunction representation \cite{ReiGam,rpanucl} or in terms of
explicit $1ph$ matrix elements. The latter reads
\begin{subequations}
\begin{eqnarray}
  \langle[\hat{X}_k,\hat{P}_{k'}]\rangle
  =2\sum_{ph}(p|\hat{X}_k|h)(h|\hat{P}_{k'}|p) ,
\\
  \langle[\hat{Y}_k,\hat{Q}_{k'}]\rangle
  =2\sum_{ph}(p|\hat{Y}_k|h)(h|\hat{Q}_{k'}|p) .
\end{eqnarray}
\end{subequations}
Alternatively,
the strength constants can be determined from the requirement
\begin{subequations}
\begin{eqnarray}
  \langle[\hat{P}_k,[\hat{V}_{\rm res},\hat{P}_{k'}]]\rangle
  &=&
  \langle[\hat{P}_k,[\hat{V}_{\rm res}^{\rm(sep)},\hat{P}_{k'}]]\rangle
  ,
  \label{eq:DCP}
\\
  \langle[\hat{Q}_k,[\hat{V}_{\rm res},\hat{Q}_{k'}]]\rangle
  &=&
  \langle[\hat{Q}_k,[\hat{V}_{\rm res}^{\rm(sep)},\hat{Q}_{k'}]]\rangle
 .
\label{eq:DCQ}
\end{eqnarray}
\end{subequations}
Substituting (\ref{eq:vres}) and (\ref{eq:Xgen}) into (\ref{eq:DCP})
yields
\begin{equation}\label{eq:PX}
  \langle[\hat{P}_k,{\hat X}_{k'}]\rangle
  =\sum_{\tilde{k}\tilde{k}'}
  \langle[\hat{P}_k,{\hat X}_{\tilde{k}}]\rangle
   \kappa_{\tilde{k}\tilde{k}'}
  \langle[{\hat X}_{\tilde{k}'},\hat{P}_{k'}]\rangle
\end{equation}
and finally Eq. (\ref{eq:k_kk'}). 
A similar procedure applies to the
double commutator (\ref{eq:DCQ}).  For a physical interpretation of the
requirements (\ref{eq:DCP}) and (\ref{eq:DCQ}) mind that the similar
double commutators but with the full Hamiltonian (instead
of the residual interaction) correspond to $m_3$ and $m_1$ sum rules,
respectively, and so represent the spring and inertia parameters
\cite{ReiBraGen} in the basis of collective generators $\hat{Q}_k$
and $\hat{P}_k$. The condition  (\ref{eq:sepstreng}) means that
collective moments $m_3$ and $m_1$ are exactly reproduced.

The basic result of this section are Eqs. (\ref{eq:sepform})
and  (\ref{eq:sepstreng})
for the separable operators and strength constants, respectively.
It is worth noting that these results can be also obtained 
\cite{Maiori} 
using the concept of nuclear self-consistency between the 
single-particle potential and density \cite{Row}. 
More detailed formulation of the basic double commutators
in terms of the SHF functional will be given in section
\ref{sec:connDFT} and appendix \ref{sec:SHF}.  The links to the
vibrating potential model (VPM) \cite{Row,BM} are outlined in appendix
\ref{sec:vpm}.  The choice of the generators $\hat{Q}_{k}$ and
$\hat{P}_{k}$ is yet open. An intuitive and pragmatic guess will be
presented and discussed in sections \ref{sec:choice} and
\ref{sec:discussion}.

\subsection{Conservation laws}

The forms (\ref{eq:sepform}) are able to reproduce given conservation
laws. Assume a symmetry mode whose generator $\hat{P}_{\rm sym}$ obeys
by definition $[\hat{H},\hat{P}_{\rm sym}]=0$. We simply have to
include this mode into the set of
generators, i.e.  $\hat{P}_{\rm
sym}\in \{\hat{P}_k\}$ together with its complement $\hat{Q}_{\rm sym}$
given by $[\hat{H},\hat{Q}_{\rm sym}]=-i\hat{P}_{\rm sym}$.
One can prove that the separable Hamiltonian (\ref{eq:hres}) does also
fulfill
\begin{equation}
  [\hat{H}_{\rm SRPA},\hat{P}_{\rm sym}]
  =0 .
\label{eq:conserv}
\end{equation}
From (\ref{eq:conserv}) we have
$$
  [\hat{h}_0,\hat{P}_{\rm sym}]
  =
  - [\hat{V}_{\rm res},\hat{P}_{\rm sym}]
  =
  -\hat{X}_{\rm sym} .
$$
Then
\begin{eqnarray*}
  [\hat{H}_{\rm SRPA},\hat{P}_{\rm sym}]
  &=&
  [\hat{h}_0,\hat{P}_{\rm sym}]
  + \sum_{k,k'}
  \hat{X}_k\kappa_{kk'}\langle[\hat{X}_{k'},\hat{P}_{\rm sym}]\rangle
\\
  &=&
  -\hat{X}_{\rm sym}
  + \sum_{k}
  \hat{X}_k\underbrace{
  \sum_{k'}\kappa_{kk'}\kappa^{-1}_{k'{\rm sym}}}_{
                       \delta_{k,{\rm sym}}}
\\
  &=&
  -\hat{X}_{\rm sym}
  +
  \hat{X}_{\rm sym}
\end{eqnarray*}
which obviously yields Eq. (\ref{eq:conserv}).
Thus all symmetry modes are recovered if properly included in the ansatz.

It is worth noting that in the case of a symmetry mode
the operator of the residual interaction is
\begin{equation}
\hat{X}_{\rm sym}=-[\hat{h}_0,\hat{P}_{\rm sym}] ,
\label{eq:Pyat_op}
\end{equation}
i.e. can be also presented as the commutator with the 
single-particle Hamiltonian.
In this case, the residual interaction serves to restore
the symmetries violated in the mean field. The similar forces were
proposed in \cite{Pyatov} to restore translational
and rotational invariance.

\subsection{Response functions}

One way to solve the SRPA equations is to compute directly
a desired strength function
\begin{eqnarray}
\label{eq:sf}
  S_D(\omega)
  &=&
  \sum_N |\langle \Psi_N |\hat{D}|\Psi_0\rangle |^2
  (\delta (\omega -\omega_N)+\delta (\omega +\omega_N))
  \nonumber\\
  &=&
  -\frac{1}{\pi}\Im\left\{\langle
  [\hat{D}, \frac{1}{\omega+i\Gamma-{\cal L}}
   \hat{D}]
   \rangle \right\}
\end{eqnarray}
where $\hat{D}$ is the operator of interest,
$\Psi_N$ and $\omega_N$ are the wave function and energy of
$N$-th RPA state,
${\cal L}$ is the Liouvillian for small amplitude excitations.
The latter is a shorthand for
\begin{equation}
  {\cal L}\hat{A}
  =
  [\hat{H},\hat{A}]_{ph}
\end{equation}
where $\hat{A}$ is any $1ph$ operator. Similarly, we introduce the
unperturbed Liouvillian ${\cal L}_0$ and the Liouvillian for the
residual interaction ${\cal V}_{\rm res}$ by
$$
  {\cal L}_0\hat{A}=[\hat{h}_0,\hat{A}] ,
\quad  %,\quad
  {\cal V}_{\rm res}\hat{A}=[\hat{V}_{\rm res},\hat{A}]_{ph}
  %\quad
  .
$$
For a compact formulation,
we collect the operators $\hat{D}$,
$\{\hat{X}_k, k=1,...,N_{\rm sep}\}$ and
$\{\hat{Y}_k, k=1,...,N_{\rm sep}\}$  into the set
$\{\hat{Z}_k, k=0,...,2N_{\rm sep}\}$ with
$\hat{X}_0=\hat{D}$.
The matrices of the strength constants are joined into
the matrix $\mu$ of the rank $2N_{\rm sep}+1$:
\begin{equation}
  \mu
  =
  \left(
  \begin{array}{ccc}
   1 & 0 & 0 \\
   0 & \kappa & 0 \\
   0 & 0 & \eta \\
  \end{array}
  \right) .
\end{equation}
Then we define generalized response functions
\begin{eqnarray}
  R_{kk'}(\omega)
  &=&
  \langle
  [\hat{Z}_k, \frac{1}{\omega+i\Gamma-{\cal L}}\hat{Z}_{k'}]
  \rangle ,
\\
  R_{kk'}^{(0)}(\omega)
  &=&
  \langle
  [\hat{Z}_k, \frac{1}{\omega+i\Gamma-{\cal L}_0}\hat{Z}_{k'}]
  \rangle
 \label{eq:R0kk} 
  .
\end{eqnarray}
Note that the free response function can be written easily 
in terms of single-particle states $p$ and $h$ as
\begin{eqnarray}
  R_{kk'}^{(0)}(\omega)
  &=&
  \sum_{ph}
  (p|\hat{Z}_k|h)(h|\hat{Z}_{k'}|p)
\nonumber\\
  &&
   \qquad\left\{
   \frac{1}{\omega\!+\!i\Gamma-\varepsilon_{ph}}
   -
   \frac{\Pi_k\Pi_{k'}}{\omega\!+\!i\Gamma+\varepsilon_{ph}}
      \right\} .
\label{eq:R0detail}
\end{eqnarray}
where $\varepsilon_{ph}$ is the energy of the 1ph-configuration
and $\Pi_k$ is the time-parity of the related operator $\hat{Z}_k$.

A relation between full and free responses is established
through the following steps:
\begin{eqnarray}
  &R_{kk'}&
  =
  \langle
  [\hat{Z}_k, \frac{1}
  {\omega+i\Gamma-{\cal L}_0-{\cal V}_{\rm res}^{\rm(sep)}}
  \hat{Z}_{k'}]\rangle
\\
  &=&
  \sum_{n=0}^\infty
  \langle
  [\hat{Z}_k, \frac{1}{\omega+i\Gamma-{\cal L}_0}
         \left({\cal V}_{\rm res}^{\rm(sep)}
               \frac{1}{\omega+i\Gamma-{\cal L}_0}\right)^n
  \!\hat{Z}_{k'}]
  \rangle
\nonumber\\
  &=&
  R_{kk'}^{(0)}
  + \sum_{k_1k_2}R_{kk_1}^{(0)}
    \langle [\hat{Z}_{k_1},[{\cal V}_{\rm res}^{\rm(sep)},
          \hat{Z}_{k_2}]]\rangle R_{k_2k'}^{(0)} + ...
\nonumber\\
  &=&
  R_{kk'}^{(0)}
  +
  \sum_{k_1k_2}R_{kk_1}^{(0)}\mu_{k_1k_2}R_{k_2k'}^{(0)}
  +  ...
 \quad . 
\nonumber
\end{eqnarray}
This can be recollected to the final result
\begin{equation}
  R_{kk'}(\omega)
  =
  \left(R^{(0)}(\omega)\frac{1}{1-\mu R^{(0)}(\omega)}\right)_{kk'}
  .
\end{equation}
Note that this operator relation deals with
matrices of the rank $2(2N_{\rm sep}\!+\!1)$ 
(the additional coefficient 2 arises due to isospin).
The strength function is finally
\begin{equation}
  S_D(\omega)
  =
  -\frac{1}{\pi}\Im\left\{R_{00}(\omega)\right\}
  .
\end{equation}

The calculation of the free response function requires some caution
with respect to time parity, see Eq. (\ref{eq:R0kk}).
The denominator in $R_{kk'}^{(0)}$ mixes time parities and one has
to disentangle the pieces
\begin{eqnarray*}
  \frac{1}{\omega+i\Gamma-{\cal L}_0}
  &=&
  {\cal R}_+^{(0)}(\omega)
  +
  {\cal R}_-^{(0)}(\omega)
  %\quad
  ,
\\
  {\cal R}_+^{(0)}(\omega)
  &=&
  \frac{{\cal L}_0}{(\omega+i\Gamma )^2-{\cal L}_0^2} ,
\\
  {\cal R}_-^{(0)}(\omega)
  &=&
  \frac{\omega+i\Gamma}{(\omega+i\Gamma )^2-{\cal L}_0^2}
\end{eqnarray*}
where
${\cal R}_+^{(0)}$ and ${\cal R}_-^{(0)}$ connect operators 
of the same and opposite parity, respectively.
One thus has to use ${\cal R}_+^{(0)}$ for
$\langle [\hat{X},{\cal R}^{(0)}\hat{X}]\rangle$,
$\langle [\hat{Y},{\cal R}^{(0)}\hat{Y}]\rangle$,
$\langle [\hat{X},{\cal R}^{(0)}\hat{D}]\rangle$,
$\langle [\hat{D},{\cal R}^{(0)}\hat{D}]\rangle$
and ${\cal R}_-^{(0)}$ for
$\langle [\hat{X},{\cal R}^{(0)}\hat{Y}]\rangle$,
$\langle [\hat{D},{\cal R}^{(0)}\hat{Y}]\rangle$.
The mutually other combinations vanish.

\subsection{SRPA as eigenvalue problem}

As often done in RPA, one wants to determine the detailed
eigenfrequencies and eigenstates of the response problem.
The excitation phonon operator $\hat{C}^{\dagger}_N$ for 
the mode $N$ is determined from the RPA equations
\begin{equation}
  [\hat{H},\hat{C}^{\dagger}_N]_{ph}
  =
  \omega_N\hat{C}^{\dagger}_N .
\label{eq:RPAcompact}
\end{equation}
The standard RPA scheme proceeds to recast that equations 
into the form of a matrix equation by expanding 
$\hat{C}^{\dagger}_N$ in terms of $1ph$ operators
\begin{equation}
  \hat{C}^{\dagger}_N
  =
  \sum_{ph}
  (c_{N,ph}^{(+)}\hat{a}_p^{\dagger}\hat{a}_h^{\mbox{}}
  - c_{N,ph}^{(-)}\hat{a}_h^{\dagger}\hat{a}_p^{\mbox{}}) .
\end{equation}
The involved matrices can grow huge in view of the large 
$1ph$ spaces required for sufficient convergence of the result.
The eigenvalue problem is dramatically simplified by inserting 
the separable Hamiltonian (\ref{eq:hres}).
For compact notation, we recombine again the sets
$\{\hat{X}_k,\hat{Y}_k\}
\longrightarrow\{\hat{Z}_k, k=1,...,2N_{\rm sep}\}$
and introduce the corresponding super-matrix of strengths
$(\kappa_{kk'},\eta_{kk'})\longrightarrow\mu_{kk'}$.
The RPA equations (\ref{eq:RPAcompact}) then become 
\begin{subequations}
\begin{eqnarray}
\label{eq:cplus}
  c_{N,ph}^{(+)}
  &=&
  \sum_{kk'}
  \mu_{kk'}\frac{(p|\hat{Z}_k|h)
  {\cal C}_{N,k'}}{\omega_N-\varepsilon_{ph}} ,
\\
  c_{N,ph}^{(-)}
  &=&
  -\sum_{kk'}\mu_{kk'}
  \frac{(h|\hat{Z}_k|p)
  {\cal C}_{N,k'}}{\omega_N+\varepsilon_{ph}} 
\label{eq:cminus}
\end{eqnarray}
\end{subequations}
with 
\begin{eqnarray}
\label{eq:ccoll}
  {\cal C}_{N,k}
  &=&
  \langle[\hat{Z}_{k},\hat{C}^{\dagger}_N]\rangle
 \\
  &=&
  \sum_{ph}\left\{ 
   (h|\hat{Z}_k|p)c_{N,ph}^{(+)}+ (p|\hat{Z}_k|h)c_{N,ph}^{(-)}
  \right\} .
  \nonumber
\end{eqnarray}
Eqs. (\ref{eq:cplus})-(\ref{eq:cminus}) reduce
the impressive number of unknowns $c_{N,ph}^{(\pm)}$ 
to the much smaller set of unknonws ${\cal C}_{N,k}$.
Such reduction (which is possible due to factorization 
of the residual interaction) yields an effective RPA 
matrix with much smaller rank.

As a next step, it is then convenient to transform 
${\cal C}_{N,k}$ to unknowns 
\begin{equation}
\label{eq:cnew}
{\bar{\cal C}}_{N,k}=2\sum_{k'}\mu_{kk'}{\cal C}_{N,k'}
\end{equation}
which are directly connected with the amplitudes $p_k(t)$ and 
$q_k(t)$ of  the scaling transformation (\ref{eq:scaling}).
For harmonic oscillations 
\begin{equation}
\label{eq:harm}
q_k(t)={\bar q}_k cos(\omega t), \quad
p_k(t)={\bar p}_k sin(\omega t)  
\end{equation}
and ${\bar{\cal C}}_{N,k}={\bar q}_k$ and ${\bar p}_k$ for 
time-even
and time-odd cases, respectively. The unknowns (\ref{eq:cnew}) 
regulate self-consistently contributions of the scaling 
operators $\hat{Q}_k$ and $\hat{P}_k$ into N-th RPA state.
Every RPA state has it own set ${\bar{\cal C}}_{N,k}$ 
(${\bar q}_k$ and ${\bar p}_k$) amplitudes.

Finally, after standard algebraic steps, we obtain the 
SRPA equation for
${\bar{\cal C}}_{N,k}$
\begin{equation}
\label{eq:seqeq}
  \sum_{k'}d_{kk'}(\omega_N){\bar{\cal C}}_{N,k'}=0
\end{equation}
with
\begin{equation}
\label{eq:S2}
  d_{kk'} (\omega_N) = 2\sum_{ph}
  \frac{(p|\hat{Z}_k|h)(h|\hat{Z}_{k'}|p)\varepsilon_{ph}}
  {\varepsilon_{ph}^2-\omega_N^2}-\mu_{kk'}^{-1} .
\end{equation}
Eq. (\ref{eq:seqeq}) has non-trivial solutions if
\begin{equation}
\label{eq:diseq}
   \mbox{det}\left\{d_{kk'}(\omega_N)\right\}=0 
\end{equation}
which yields the RPA spectrum $\omega_N$.
Eq. (\ref{eq:seqeq}) is basically a matrix of the rank 
$4N_{\rm sep}$ (after duplication of the rank because of the 
isospin) which is much smaller than the necessary number of 
$1ph$ states.  Typically, we can deal with $N_{\rm sep}=1-4$ 
(see section \ref{sec:discussion}) as opposed to the rank
of a conventional RPA matrix which often amounts to about 
$10^3-10^6$.

The normalization condition for RPA states is
\begin{equation}
  \langle (\hat{C}_N^{\mbox{}},\hat{C}_N^{\dagger})\rangle
  =1
\end{equation}
which amounts to
\begin{eqnarray}
\label{eq:norm_se}
&&\sum_{ph}(|c_{N,ph}^{(+)}|^2-|c_{N,ph}^{(-)}|^2)
\\
 &=&
  \sum_{kk'}{\bar{\cal C}}_{N,k}{\bar{\cal C}}_{N,k'}
\sum_{ph}\frac{<p\vert \hat{Z}_{k} \vert h>
    <p\vert \hat{Z}_{k'} \vert h>\varepsilon _{ph}\omega_N}
    {(\varepsilon _{ph}^2 - \omega^2_j)^2}
\nonumber\\
&=&\frac{1}{4}
\sum_{kk'}{\bar{\cal C}}_{N,k}{\bar{\cal C}}_{N,k'}
\frac{\partial}{\partial\omega_N}d_{kk'}(\omega_N)=2 .
\nonumber
\end{eqnarray}
The last line of (\ref{eq:norm_se}) represents the 
normalization condition
in terms of derivatives of the RPA matrix elements. 

The reduced probability of $E\lambda$-transition from 
the ground state to the RPA state $\omega_N$ can be written as 
\begin{equation}
B(E\lambda, gr \rightarrow \omega_N)=
\frac{\sum_{kk'}D_{kk'}(\omega_N)A_k(\omega_N)A_{k'}(\omega_N)}
     { \frac{\partial}{\partial\omega_N}
        \mbox{det}\left\{d_{kk'}(\omega_N)\right\}} 
\end{equation}
where $D_{kk'}(\omega_N)$ is the algebraic supplement of the 
matrix element $d_{kk'}(\omega_N)$,
\begin{equation}
A_k(\omega_N)=\sum_{ph}\frac{\varepsilon_{ph}
  (p|\hat{Z}_k|h)(h|\hat{f}_{\lambda}|p)}
{\varepsilon_{ph}^2-\omega_N^2} 
\end{equation}
and $(h|\hat{f}_{\lambda}|p)$ is the 1ph matrix element 
of $E\lambda$-transition. 

\subsection{Connection with density functional theory}
\label{sec:connDFT}

Self-consistent models are often formulated in terms of a local
energy-density functional rather than through an explicit Hamiltonian
(see, e.g., the nuclear Skyrme-Hartree-Fock (SHF) method
\cite{SHFrev} or the Kohn-Sham scheme in electron systems
\cite{KohSha,DreGro}). So we will work out 
the above SRPA scheme in terms of a given energy functional
$$
  E=E(\{J_\alpha({\vec r})\})
$$
where
$$
  J_\alpha({\vec r})
  =
  \langle\Psi|\hat{J}_\alpha({\vec r})|\Psi\rangle
$$
constitute a set of local densities and currents associated with
the functional. A simple example is the situation in atoms 
and atomic clusters where we deal only with the local density 
$\rho(\vec{r})=\sum_n|\varphi_n(\vec{r})|^2$ of valence electrons.
The nuclear case is much more involved. The typical set for Skyrme
functional includes, for both protons and neutrons, the local 
density $\rho$, kinetic density $\tau$, current ${\bf j}$, 
spin-orbit density $\vec{J}$, and spin density ${\bf \sigma}$
i.e.
\begin{equation}
\label{eq:set}
  J_\alpha
  \in
  \{\rho_p,\rho_n,\tau_p,\tau_n,\vec{J}_p,\vec{J}_n,
    \vec{j}_p,\vec{j}_n,\vec \sigma_p,\vec \sigma_n\} ,
\end{equation}
see in Appendix \ref{sec:SHF} the detailed expressions.

The mean field Hamiltonian corresponding to the given energy 
functional is then determined as
\begin{equation}
  \hat{h}(\vec{r})
  =
  \int d\vec{r}\,\sum_{\alpha}{\hat J}_{\alpha}(\vec{r})
   \frac{\delta E}{\delta J_{\alpha}(\vec{r})} .
\label{eq:h}
\end{equation}
In the linear regime of small-amplitude deformation generated 
by a $1ph$ operator $\hat{G}$, see Eq. (\ref{eq:scaling}), 
the densities can be decomposed as
\begin{equation}
  J_{\alpha}({\vec r},t)
  =
  {\bar J}_{\alpha}({\vec r})
  +
  {J}^G_{\alpha}({\vec r},t)
\label{eq:delJ}
\end{equation}
where ${\bar J}_{\alpha}({\vec r})$ is the static ground-state 
density and
\begin{eqnarray}
  {J}^G_{\alpha}({\vec r},t)
  &=&
  \langle\Psi(t)|{\hat J}_{\alpha}(\vec{r})|\Psi(t)\rangle -
  \langle\Psi_0|{\hat J}_{\alpha}(\vec{r})|\Psi_0\rangle
\nonumber\\
  &=&
  \langle[{\hat J}_{\alpha}(\vec{r}),\hat{G}(\vec{r},t)]\rangle
\label{eq:defJG}
\end{eqnarray}
accounts for the small change through the deformation. Inserting 
(\ref{eq:delJ}) into the mean-field Hamiltonian (\ref{eq:h}) and
expanding the latter in orders of
$\hat{G}$, we get, in the first order, the response Hamiltonian
$[\hat{V}_{\rm res},\hat{G}]_{ph}^{\mbox{}}$
in terms of the density functional
\begin{equation}
  [\hat{V}_{\rm res},\hat{G}]_{ph}^{\mbox{}}
 =
 \sum_{\alpha\beta}\!\int\!\! 
 d{\vec r}d{\vec r}'{\hat J}_{\alpha}({\vec r})
 \frac{\delta^2 E}{\delta J_{\alpha}({\vec r})\delta J_{\beta}({\vec r}')}
 {J}^G_{\beta}({\vec r}',t) .
\label{eq:hres2}
\end{equation}

We now decompose the scaling state as in Eq. (\ref{eq:scaling}),
which yields the analogous decomposition for the variation of density
\begin{equation}
  {J}^G_{\alpha}({\vec r},t)
  =
  -i\sum_k\left(p_k(t){J}^{Q_k}_{\alpha}({\vec r})
              -q_k(t){J}^{P_k}_{\alpha}({\vec r})\right)
\end{equation}
where, similar to Eq. (\ref{eq:defJG}),
\begin{subequations}
\begin{eqnarray}
  J^{Q_k}_{\alpha}({\vec r})
  &=&
  \langle[{\hat J}_{\alpha}({\vec r}),\hat{Q}_{k}({\vec r})]\rangle ,
  \\
  {J}^{P_k}_{\alpha}({\vec r})
  &=&
  \langle[{\hat J}_{\alpha}({\vec r}),\hat{P}_{k}({\vec r})]\rangle .
\end{eqnarray}
\end{subequations}
Every step proceeds much similar as for the general $\hat{G}$ 
because we are in the linear regime. Finally, we can specify 
the separable operators (\ref{eq:Xgen}) and (\ref{eq:Ygen}) as
\begin{subequations}
\begin{eqnarray}
  \hat{X}_k(\vec{r})
  &\!=\!&
 \!\sum_{\alpha\beta}\!\int\!\! 
 d{\vec r}d{\vec r}'{\hat J}_{\alpha}({\vec r})
 \frac{\delta^2 E}{\delta J_{\alpha}({\vec r})
 \delta J_{\beta}({\vec r}')}
 {J}^{P_k}_{\beta}({\vec r}') ,
\label{eq:x}
\\
  \hat{Y}_{k}(\vec{r})
  &\!=\!&
 \!\sum_{\alpha\beta}\!\int\!\! 
 d{\vec r}d{\vec r}'{\hat J}_{\alpha}({\vec r})
 \frac{\delta^2 E}{\delta J_{\alpha}({\vec r})
 \delta J_{\beta}({\vec r}')}
 {J}^{Q_k}_{\beta}({\vec r}') .
\label{eq:y}
\end{eqnarray}
\end{subequations}
and the strength matrices (\ref{eq:k_kk'})
and (\ref{eq:e_kk'}) as
\begin{subequations}
\begin{eqnarray}
\label{eq:kap_dft}
  \kappa_{kk'}^{-1}
  &\!=\!&
 \!\!\sum_{\alpha\beta}\!\!\int\!\!
 d{\vec r}d{\vec r}'{J}^{P_k}_{\alpha}({\vec r})
 \frac{\delta^2 E}{\delta J_{\alpha}({\vec r})
 \delta J_{\beta}({\vec r}')}
 {J}^{P_{k'}}_{\beta}({\vec r}')
 \,,
\\
\label{eq:eta_dft}
  \eta_{kk'}^{-1}
  &\!=\!&
 \!\!\sum_{\alpha\beta}\!\!\int\!\! 
 d{\vec r}d{\vec r}'{J}^{Q_k}_{\alpha}({\vec r})
 \frac{\delta^2 E}{\delta J_{\alpha}({\vec r})
 \delta J_{\beta}({\vec r}')}
 {J}^{Q_{k'}}_{\beta}({\vec r}')
 \,.
\end{eqnarray}
\end{subequations}
The latter equations demonstrate the symmetry of the 
strength matrices $\kappa_{kk'}$ and $\eta_{kk'}$.
Note the subtle difference between separable operators 
and strengths. The double integrals look much the same. 
However, in Eqs. (\ref{eq:x}) and (\ref{eq:y}), they 
contain the current operators $\hat{J}_\alpha(\bf{r})$ 
and yield one-body operators as a result of the integration. 
Instead, Eqs. (\ref{eq:kap_dft}) and (\ref{eq:eta_dft}) do 
not include operators and so yield strength coefficients 
as c-numbers. For more details see Appendix B. 

\subsection{Choice of collective generators}
\label{sec:choice}

As was mentioned above, the proper choice of generating operators
$\hat{Q}_{k}({\vec r})$ is crucial to achieve good convergence of the
separable expansion (\ref{eq:vres}) with a minimal number of separable
operators. The choice is inspired by both physical and computational
arguments. It should be simple and universal in the sense that it can
be applied equally well to all modes and excitation channels. The main
idea is that the generating operators should explore different spatial
regions of the nucleus, the surface as well as the interior.  This
suggests that the leading scaling operator should have the form of the
applied external field in the long-wave approximation, for example,
$\hat{Q}^{\lambda\mu}_{k=1}({\vec r})=r^{\lambda}
(Y_{\lambda\mu}(\Omega)+\mbox{h.c.)}$. Such a choice results in
separable operators (\ref{eq:x}) and (\ref{eq:y}) 
which are most sensitive to
the surface of the system. The first excitation modes of nuclear
collective motion tend to vibrate predominantly in the surface region.
As a result, already with this one separable term we obtain a quite
good description of giant resonances. The detailed distributions
depends on a subtle interplay of surface and volume vibrations.  This
can be resolved by taking into account an interior of the nucleus. For
this aim, the radial parts with larger powers
$r^{\lambda+p}Y_{\lambda\mu}$ and spherical Bessel functions can be
used, much similar as in the local RPA
\cite{ReiBraGen,rpanucl}. This results in the shift of the maxima 
of the operators (\ref{eq:x}) and (\ref{eq:y}) to the interior. 
Exploring different conceivable combinations,
we have found a most efficient set of the generators as
\begin{equation}
  \hat{Q}_{k}({\vec r})
  =
  R_k(r)(Y_{\lambda\mu} (\Omega )+\mbox{h.c.})
  \quad,\quad
  \hat{P}_k
  =
  i[\hat{H},\hat{Q}_k]
\label{eq:scale_op}
\end{equation}
with
\begin{equation}
  R_k(r)
  =
  \left\{
  \begin{array}{ll}
   r^{\lambda }, & k\!=\!1
  \\
  j_{\lambda}(q^k_{\lambda}r), & k\!=\!2,3,4
\\
\end{array}
\right.
\label{eq:actualset}
\end{equation}
$$
  q^{k}_{\lambda} = a_k\frac{z_{\lambda}}{R_{\rm diff}} ,\quad
  a_2\!=\!0.6\;,
  a_3\!=\!0.9\;,
  a_4\!=\!1.2
$$
where $R_{\rm diff}$ is the diffraction radius of the actual nucleus
and $z_{\lambda}$ is the first root in $j_{\lambda}(z_{\lambda})=0$.  
The separable term with $k=1$ is
mainly localized at the nuclear surface while the next three terms are
localized more and more in the interior.  This simple set seems to be
a best compromise for the description of nuclear giant resonances in
light and heavy nuclei.

One may argue that one needs to explore not only different spatial
regions but also the different types of the operators, associated 
with the variety of densities and currents (e.g. kinetic, spin-orbit). 
We have checked this. And indeed, this allows a perfect fine 
tuning of the spectra. On the other hand, it adds more separable 
terms. The set (\ref{eq:scale_op})-(\ref{eq:actualset}) with
few and purely local generators is in our opinion the best compromise
between quality of the results and the expense of the calculations.

\section{Results and discussion}
\label{sec:discussion}

We test SRPA for Skyrme forces by comparison with standard full 
RPA calculations. As the test
cases we consider the isovector dipole and isoscalar quadrupole
resonances in the two doubly magic nuclei $^{40}{\rm Ca}$ and
$^{208}{\rm Pb}$. Most tests are performed with widely used Skyrme
force SkM* \cite{SkM}.  
The convergence of the SRPA expansion is found to be only slightly  
dependent on the actual force, which is demonstrated  
for the case of the recent parameterization SkI3
\cite{FloRei}. The full RPA calculations are done in a large basis
which covers the $1ph$ excitations from all hole states to all
particle states up to 30 MeV above the Fermi energy. SRPA 
easily allows to extend this basis but
 we restrict the $1ph$ summations in SRPA to precisely the
same active space as used in full RPA to have one-to-one comparison.
The comparison is done in terms of the strength function ${\cal
S}_D(\omega)$ using a smoothing parameter of $\Gamma=1\,{\rm MeV}$. 
Specifically, the Lorentz weight function was used. 
The value of the smoothing is quite realistic because 
experimental spectra are typically washed out by similar 
widths.

%%%%%%%%%%%%%%%%%%%%%%%%
%                      %     
% Place for Figure 1.  %
%                      % 
%%%%%%%%%%%%%%%%%%%%%%%%

Figure 1
%\ref{fig:fig1}  
compares SRPA and full RPA results for SkM* forces.   
Three stages of SRPA expansion with increasing number of 
separable terms are shown: only one
term $k\!=\!1$, two terms $k\!=\!1,4$, and all four terms according to
the labeling (\ref{eq:actualset}). Mind that we use in the second set
the operator with $a_4=1.2$ since then the forces cover the extremes of
both surface (k=1) and volume (k=4) sensitivity. The first stage
with only $k\!=\!1$ provides already very good description of the
isoscalar E2 resonance in both nuclei.  Even the low-energy quadrupole
mode in $^{208}{\rm Pb}$ is well reproduced. The influence of further
terms is very small.   
The isovector E1 resonance seems to be more demanding.  The first
stage correctly describes the average position and width of the
resonance but cannot fully cope with the detailed fragmentation
pattern. Here one sees clearly the systematic improvement achieved by
the further terms.  The set with four terms provides a satisfying
agreement. Further improvement is possible by adding more terms
but the extra effort seems to be unnecessary. The agreement achieved 
is already more than sufficient for any practical calculations. 
One may even be content with a smaller sets for large scale 
exploratory studies.

%%%%%%%%%%%%%%%%%%%%%%%%
%                      %     
% Place for Figure 2.  %
%                      % 
%%%%%%%%%%%%%%%%%%%%%%%%

Although the smoothing with 1 MeV provides the typical realistic
shapes of strength distributions, it is worth having a quick glance at
spectra with higher resolution.  This is done in
Fig. 2
%\ref{fig:fig2} 
drawn with smoothing $\Gamma =$0.1 MeV. The
figure provides the detailed comparison of SRPA and exact RPA results
for the most complicated cases of the isovector E1 resonance.  Though
we have not a perfect coincidence (which may be partly excused by
numerical noise in the full RPA calculations), the results are
indeed extremely close. This confirms the reliability of SRPA 
even at this finer scale.

%%%%%%%%%%%%%%%%%%%%%%%%
%                      %     
% Place for Figure 3.  %
%                      % 
%%%%%%%%%%%%%%%%%%%%%%%%

%%%%%%%%%%%%%%%%%%%%%%%%
%                      %     
% Place for Figure 3.  %
%                      % 
%%%%%%%%%%%%%%%%%%%%%%%%

Figure 3 
%\ref{fig:fig3} 
demonstrates the quality of SRPA for another Skyrme force, 
namely the parameterization SkI3.  It shows the toughest 
test case, the isovector E1 resonance in $^{208}{\rm Pb}$.  
The
full set k=1-4 is used.  Again we see a satisfying reproduction of
this "worst case". The other test cases (not shown here) perform
even better. We also checked a few other forces and always found
the same quality.

In fact, the spectrum from SkI3 looks much more realistic than that
obtained from SkM* (see lower right panel in Fig. 1.
%\ref{fig:fig1}). 
The pronounced right shoulder in SkM* is, to a large extend, 
an artefact from the restricted $1ph$ space used in
these calculations. SRPA offers the opportunity to remove that
restriction and to perform calculations in the full space of
excitations, limited only by numerical resolution of the underlying
coordinate space grid (with particle energies at least up to 
600 MeV). Figure 4
%\ref{fig:comp_spaces} 
compares SRPA results in the restricted and unrestricted spaces.  
It is obvious that the consistent inclusion of
$1ph$ states above 30 MeV serves to diminish the unnaturally high
shoulder at 16 MeV. This is a clear hint that RPA 
calculations converge
slowly with the size of phase space. The SRPA allows to
go much further in that respect.

Altogether, the tests show that the proposed set of four operators
suffices for a satisfying reproduction of full RPA. The number
"four" has to be taken with a grain of salt. We have four operators
of the type $\hat{Q}_k$ for protons and another four for neutrons.
The set is complemented by the conjugate operators
$\hat{P}_k=i[\hat{H},\hat{Q}_k]$. This amounts to 16 operators in total
and the computation of the strength function requires inversion of
$17\!\times\!17$ matrices. The full RPA calculations, on the other
hand, used matrices up to a rank of 400 and more.

\section{Conclusions}

A novel self-consistent separable approach to the random phase
approximation (SRPA) is proposed. SRPA employs a systematic expansion
of the exact residual interaction into a sum of weighted separable
terms (i.e. products of one-body operators evaluated at a Hartree
level).  For both, the weights and the separable operators, compact
analytical expressions in terms of collective generators are derived
in a self-consistent manner. The form of the separable
residual interaction is optimized by the condition that the vibrating 
mean field as generated from collective flow reproduces the exact result. 
SRPA is formulated in a general way such that it can be applied
to arbitrary energy functionals depending on various local densities
and currents.
As open point remains the choice of the relevant collective
generators. Here we exploit experience gained in nuclear fluid
dynamical models (local RPA) and use an efficient mix of surface and
volume modes.

The general SRPA scheme is specified for the particular case of the
Skyrme energy functional. The convergence of the separable expansion
and performance of SRPA for giant resonances in light and heavy nuclei
is investigated. The results are very encouraging. SRPA proves to be a
very good approximation to exact RPA and it it requires much less
numerical effort.  In particular, it allows to work with huge
particle-hole spaces, giving thus the chance for unrestricted RPA
calculations, even in nuclei with broken symmetries
%any dimensionality, i.e. for spherical, axial, and triaxial nuclei.

\section*{Acknowledgments}
 This work was partly supported by the Czech grant agency under the contract
No. 202/02/0939 (J.K.). V.O.N.  thanks the support of RFBR (00-02-17194),
Heisenberg-Landau (Germany-BLTP JINR), DFG (436RUS17/102/01) and
Votruba-Blokhintcev (Czech Republic -BLTP JINR) grants.

\appendix
\section{Comparison with VPM}
\label{sec:vpm}

\bigskip

It is interesting to compare SRPA with the Vibrating Potential 
Model (VPM) \cite{Row,BM} whose different versions are widely 
used in nuclear structure theory (see, for example,
\cite{LS,SS,Ne_PRC,Kubo,Aberg}). VPM deals with the mean field
\begin{equation}
\label{A1}
\hat{h}_0=T+U_0
\end{equation}
where $U_0$ is purely local. Moreover, $U_0$ is often a 
density-independent phenomenological potential (oscillator, 
Nilsson, Woods-Saxon, ...). The vibrations of the system with 
the mean field (\ref{A1}) are characterized by the collective 
deformations $q_{k}$ and in the framework of VPM the
vibrating potential is written as
\begin{equation}
\label{A2}
U=U_0+\delta U ,\quad
\delta U=\sum_{k} \frac{\partial U}{\partial q_{k}}\delta q_{k}
=-i\sum_{k} X_{k} \delta q_{k}
\end{equation}
where $X_{k}\!\!=\!\! i\frac{\partial U}{\partial q_{k}}$ are 
single-particle operators in the r-representation, participating 
in VPM separable residual interactions (see e.g. \cite{Aberg}). 
Usually the collective momenta $P_{k}$ are introduced
\begin{equation}
\label{A3}
P_{k}= i\frac{\partial }{\partial q_{k}} .
\end{equation}
This allows to write
\begin{eqnarray}
\label{A4}
X_{k}=i \frac{\partial U}{\partial q_{k}}
=[U,P_{k}] \quad \Rightarrow \quad   \nonumber\\
\hat{X}_{k}=\int \! d{\vec r} \hat{\rho}(\vec{r}) 
 [\hat{U},\hat{P}_{k}]
 \approx  \int \! d{\vec r} \hat{\rho}(\vec r) 
 <[\hat{U},\hat{P}_{k}]>
\end{eqnarray}
where $\hat{X}_{k}$ is the operator corresponding to $X_{k}$ 
and acting in the Fock space. VPM operators (\ref{A4}) should 
be compared with the SRPA operators (\ref{eq:Xgen}) or (\ref{eq:x}). 
For the first sight Eq. (\ref{A4}) contradicts to (\ref{eq:Xgen}). 
However, there are, in fact, more similarities than differences. 
The equivalence of (\ref{A4}) and (\ref{eq:x}) can be demonstrated 
in the case of the simple energy functional
\begin{equation}
\label{A5}
  E=
  \textstyle{\frac{1}{2}}t_0\int d{\vec r}\,\rho(\vec r)^2 .
\end{equation}
This functional gives the following mean field potential
\begin{equation}
\label{A6}
U=t_0\rho .
\end{equation}
VPM expression (\ref{A4}) gives
\begin{equation}
\label{A7}
\hat{X}_{k}= t_{0} \int \! \! d{\vec r} \hat{\rho}(\vec r)
<[\hat{\rho}(\vec r),\hat{P}_{k}]> ,
\end{equation}
while SRPA expression (\ref{eq:x}) provides
\begin{eqnarray}
\label{A8}
\hat{X}_{k} &=& \int \! d{\vec r} d{\vec r}' \hat{\rho}(\vec r)
\frac{\delta^2 E}{\delta \rho(\vec r)\delta \rho(\vec{r}')}
<[\hat{\rho}(\vec{r}'),\hat{P}_{k}]> \nonumber\\
&=&  t_{0}\int \! \! d{\vec r} \hat{\rho}(\vec r)
<[\hat{\rho}(\vec r),\hat{P}_{k}]>.
\end{eqnarray}
Thus there is no difference between VPM and SRPA for simple
two-body forces following from the functional (\ref{A5}). 
Differences come up with non-linear terms in the 
mean-field potential. From this point of view SRPA seems
to be more general.

\section{Details of SHF}
\label{sec:SHF}

A most widely used energy functional for nuclear structure
calculations is the Skyrme functional. It was originally proposed in
\cite{skyrme} and made its breakthrough with the first fine-tuning of
\cite{VauBri}. Since then it has been steadily further developed and
constitutes today a very reliable model for nuclear structure and
excitations.  The functional involves the set of local densities and
currents (\ref{eq:set}) which read in general
\begin{eqnarray*}
  \rho_q({\vec r})
  &=&
  \sum_{n\in q}
  \varphi^*_n({\vec r})\varphi_n^{\mbox{}}({\vec r}) ,
\\
  \tau_q({\vec r})
  &=&
  \sum_{n\in q}
  \nabla\varphi^*_n({\vec r})\!\cdot\!
  \nabla\varphi_n^{\mbox{}}({\vec r}) ,
\\
  \vec{J}_q({\vec r})
  &=&
  -i\sum_{n\in q}
  \varphi^*_n({\vec r})(\nabla\times
  \hat{\vec{\sigma}})\varphi_n^{\mbox{}}({\vec r}) ,
\\
  \vec{j}_q({\vec r})
  &=&
  -\frac{i}{2}\sum_{n\in q}
  \left[
  \varphi^*_n({\vec r})\nabla\varphi_n^{\mbox{}}({\vec r})
  -
  \nabla\varphi^*_n({\vec r})\varphi_n^{\mbox{}}({\vec r})
  \right] ,
\\
  \vec{\sigma}_q({\vec r})
  &=&
  \sum_{n\in q}
  \varphi^*_n({\vec r})\hat{\vec{\sigma}}
  \varphi_n^{\mbox{}}({\vec r}) .
\end{eqnarray*}
The associated operators are
\begin{eqnarray*}
  \hat{\rho}_q(\vec{r})
  &=&
  \hat{\Pi}_q\delta(\hat{\vec{r}}'-\vec{r}) ,
\\
  \hat{\tau}_q(\vec{r})
  &=&
  \hat{\Pi}_q\hat{\overleftarrow{\vec p}}
  \delta(\hat{\vec{r}}'-\vec{r})\hat{\vec{p}} ,
\\
  \hat{\vec{J}}_q(\vec{r})
  &=&
  \hat{\Pi}_q\delta(\hat{\vec{r}}'-
  \vec{r})\hat{\vec{p}}\!\times\!\hat{\vec{\sigma}} ,
\\
  \hat{\vec{j}}_q(\vec{r})
  &=&
  \frac{1}{2}\hat{\Pi}_q\left\{\hat{\vec{p}},
  \delta(\hat{\vec{r}}'-\vec{r})\right\} ,
\\
  \hat{\vec{\sigma}}_q(\vec{r})
  &=&
  \hat{\Pi}_q\delta(\hat{\vec{r}}'-\vec{r})\hat{\vec{\sigma}}
\end{eqnarray*}
where $\hat{\Pi}_q$ is the isospin projector to $q\in\{p,n\}$.

We use here the Skyrme functional in the form \cite{rpanucl}
\begin{eqnarray}
   {E} & = & \int d{\vec r}\left({\cal E}_{\rm kin}
 +{\cal E}_{\rm Sk}(\rho_q,\tau_q,
   \vec{\sigma}_q,\vec{j}_q,\vec{J}_q)
                           \right) 
			   \nonumber\\
             & &+{E}_{\rm C}(\rho_p)
               -E_{\rm cm} ,
\\
   {\cal E}_{\rm kin} &= &  \frac{\hbar^2}{2m} \tau ,
 \label{Ekin}
 \\  
%\end{eqnarray}
%\begin{eqnarray}
 {E}_{\rm C} 
 & = & \frac{e^2}{2}   
 \int d{\vec r} \, d{\vec r}' \rho_p(\vec{r})
              \frac{1}{|\vec{r}-\vec{r}'|} \rho_p(\vec{r}')
\nonumber\\
     & &
         -\frac{3}{4} e^2(\frac{3}{\pi})^\frac{1}{3} \int d^3r
                          [ \rho_p(\vec{r})]^\frac{4}{3} ,
\label{Ecoul}
\end{eqnarray}
\begin{widetext}
\begin{eqnarray}   
   {\cal E}_{\rm Sk} &= &
                 +\frac{b_0}{2}  \rho^2
                  -\frac{b'_0}{2} \sum_q \rho_q^2
          + \frac{b_3}{3}  \rho^{\alpha +2}
          - \frac{b'_3}{3} \rho^\alpha   \sum_q\rho_q^2
%            \nonumber \\
%     & &
     +  b_1 (\rho \tau - \vec{j}^2)
     - b'_1 \sum_q (\rho_q \tau_q - \vec{j}_q^2)
     \nonumber \\     
	    & &
     -\frac{b_2}{2} \rho \Delta \rho
     +\frac{b'_2}{2} \sum_q \rho_q \Delta \rho_q
%\nonumber\\
%     & &
     - b_4 \left( \rho\nabla\cdot\vec{{J}}
      + \vec{\sigma} \cdot (\nabla \times \vec{j})\right)
\nonumber\\
     & &
     -b'_4 \sum_q [ \rho_q(\nabla\cdot\vec{J}_q)
              + \vec{\sigma}_q \cdot (\nabla \times \vec{j}_q)  ]
%\nonumber\\
%     & &
      - \tilde{b}_4\vec{J}^2 - \tilde{b}'_4\sum_q\vec{J}_q^2 .
 \label{Esky}       
\end{eqnarray}
\end{widetext}      
The sum $\sum_q$ runs over proton or neutron species. A density 
without index means total density, e.g. $\rho=\rho_p+\rho_n$ 
and similarly for the others. The parameters $b_i$ and $b'_i$ 
used in the above definition are chosen to give a most compact 
formulation of the energy functional, the corresponding 
mean-field Hamiltonian and residual interaction . They
are related to the standard Skyrme parameters $t_i$ and $x_i$
\cite{VauBri,SHFrev} by
\begin{equation}
  \begin{array}{rcl}
   b_0 & = &   t_0 (1+\frac{1}{2} x_0) ,
 \\[1pt]
   b'_0& = &   t_0 (\frac{1}{2}+x_0) ,
 \\[1pt]
   b_1 & = &
     \frac{1}{4} \left[ t_1 (1+\frac{1}{2} x_1)+
      t_2 (1+\frac{1}{2} x_2)\right] ,
 \\[1pt]
   b'_1& = &  \frac{1}{4} \left[ t_1 (\frac{1}{2}+x_1)
              -t_2 (\frac{1}{2}+x_2)
              \right] ,
 \\[1pt]
   b_2 & = &
     \frac{1}{8} \left[ 3t_1 (1+\frac{1}{2} x_1)-
     t_2 (1+\frac{1}{2} x_2)\right] ,
 \\[1pt]
   b'_2& = &  \frac{1}{8} \left[ 3t_1 
   (\frac{1}{2}+x_1)+t_2 (\frac{1}{2}+x_2)
              \right] ,
 \\[1pt]
   b_3 & = &  \frac{1}{4} t_3 (1+\frac{1}{2} x_3) ,
 \\[1pt]
   b'_3& = &  \frac{1}{4} t_3 (\frac{1}{2}+x_3) ,
  \\[1pt]
   b_4 & = & b'_4 = \frac{1}{2} t_4 .
  \end{array}
\label{bdef}
\end{equation}
Various versions exist for the spin-orbit term. The 
$\vec{J}^2$ term is sometimes considered, sometimes not. 
Thus one has the choice
\begin{equation}
  \begin{array}{rcl}
    \tilde{b}_4 &=& -\frac{1}{16}(t_1x_1+t_2x_2) ,
  \\[1pt]
    \tilde{b}'_4 &=& +\frac{1}{16}(t_1-t_2)
  \end{array}
\end{equation}
or
\begin{equation}
    \tilde{b}_4 = 0 ,\quad\quad
    \tilde{b}'_4 = 0 .
\end{equation}
The $\rho\nabla\cdot\vec{{J}}$ term is always taken into
account. Two options exist here. Conventional Skyrme forces use
$b'_4=b_4$ while a recent extension allows for more flexible isospin
dependence considering $b_4$ and $b'_4$ as independent parameters
\cite{FloRei}. The center-of-mass correction is
$E_{\rm cm}=\langle\hat{P}_{\rm cm}^2\rangle/2mA$, often approximated
by its diagonal terms
$E_{\rm cm}=\langle\sum_i\hat{p}_i^2\rangle/2mA$. The full correction
is usually subtracted a posteriori and the approximate correction is
expectation value of a one-body operator. In any case, there is no
contribution to the residual interaction for RPA.

First variation with respect to $J_\alpha$ yields the mean-field
Hamiltonian. It can be found in many publications, e.g. 
\cite{rpanucl}.
The step of interest is the second variation yielding the
residual interaction. We obtain here
\begin{widetext}
\begin{eqnarray*}
  \frac{\delta^2 E}{\delta\rho_q(\vec{r})\delta\rho_{q'}(\vec{r}')}
  &=&
  \Big(
    b_0+b_3\frac{(\alpha\!+\!1)(\alpha\!+\!2)}{3}\rho^\alpha
%\nonumber\\
%  &-&
%\;\;
    -b'_3\frac{\alpha(\alpha\!-\!1)}{3}\rho^{\alpha-2}\sum_q\rho_q^2
%\nonumber\\
%  &-&
%\;\;
   -b'_3\frac{2\alpha}{3}\rho^{\alpha-1}
   -b_2\Delta  \Big)\delta(\vec{r}-\vec{r}')
\nonumber\\
   & &
  -\Big(
      b'_0+b'_3\frac{2}{3}\rho^\alpha
    +b'_3\frac{2\alpha}{3}\rho^{\alpha-1}\rho_q
%\nonumber\\
%   &+&
%\;\;
   +b'_2\Delta
  \Big)\delta_{qq'}\delta(\vec{r}-\vec{r}')
  \nonumber\\
   & &
   +\delta_{qp}\delta_{pq'}\Big(\frac{e^2}{|\vec{r}-\vec{r}'|}
%\nonumber\\
%   &-&
%\;\;
   -\frac{1}{3} e^2(\frac{3}{\pi})^\frac{1}{3}\rho_p^{-2/3}
    \delta(\vec{r}-\vec{r}')\Big) ,
\end{eqnarray*}
\end{widetext}
\begin{eqnarray*}
  \frac{\delta^2 E}{\delta\rho_q(\vec{r})
  \delta\tau_{q'}(\vec{r}')}
  &=&
   b_1\delta(\vec{r}-\vec{r}')-b'_1
   \delta(\vec{r}-\vec{r}')\delta_{qq'} ,
\\
  \frac{\delta^2 E}{\delta\rho_q(\vec{r})
  \delta\vec{J}_{q'}(\vec{r}')}
  &=&
   -\Big(b_4-b'_4\delta_{qq'}\Big)\vec{\nabla}
   \delta(\vec{r}-\vec{r}') ,
\\
  \frac{\delta^2 E}{\delta{\vec J}_{\nu q}(\vec{r})
  \delta{\vec J}_{\nu'q'}(\vec{r}')}
  &=&
   -\Big(\tilde{b}_4
   +\tilde{b}'_4\delta_{qq'}\Big)
   \delta(\vec{r}-\vec{r}')\delta_{\nu\nu'} ,
\\
  \frac{\delta^2 E}{\delta{j}_{\nu q}(\vec{r})
  \delta{j}_{\nu'q'}(\vec{r}')}
  &=&
   -\Big(2{b}_1
   -2{b}'_1\delta_{qq'}\Big)\delta(\vec{r}-\vec{r}')
   \delta_{\nu\nu'} ,
\\
  \frac{\delta^2 E}{\delta{j}_{\nu q}(\vec{r})
  \delta{\sigma}_{\nu'q'}(\vec{r}')}
  &=&
   -\Big({b}_4
   +{b}'_4\delta_{qq'}\Big)
   {\nabla}_\mu\!\!\times\!\delta(\vec{r}-\vec{r}')
   \epsilon_{\nu\nu'\mu}
\end{eqnarray*}
where $\nu,\nu'\in\{x,y,z\}$ labels the spatial vector 
components and
$\epsilon_{\nu\nu'\mu}$ is the antisymmetric tensor.
All combinations of derivatives not listed above vanish.

It is instructive to write the response currents in detail. 
The non-vanishing contributions are
\begin{eqnarray*}
  \rho_q^{P_k}(\vec{r})
  &=&
  <[\hat{\rho}_q,\hat{P}_k]>
  \\
  &=&
  \sum_{n\in q}\left(
  \varphi^*_n\hat{P}_k\varphi_n^{\mbox{}}
  -\hat{P}_k\varphi^*_n\varphi_n^{\mbox{}}
  \right) ,
\\
  \tau_q^{P_k}(\vec{r})
  &=&
  <[\hat{\tau}_q,\hat{P}_k]>
  \\
  &=&
  \sum_{n\in q}\left(
  \nabla\varphi^*_n\!\cdot\!\nabla\hat{P}_k\varphi_n^{\mbox{}}
  -
  \nabla\hat{P}_k\varphi^*_n\!\cdot\!\nabla\varphi_n^{\mbox{}}
  \right) ,
\\
  \vec{J}_q^{P_k}(\vec{r})
  &=&
  <[\hat{\vec{J}}_q,\hat{P}_k]>
  \\
  &=&
  -i\sum_{n\in q}\left(
  \varphi^*_n(\nabla\times\hat{\vec{\sigma}})
  \hat{P}_k\varphi_n^{\mbox{}}
  -
  \hat{P}_k\varphi^*_n(\nabla\times
  \hat{\vec{\sigma}})\varphi_n^{\mbox{}}
  \right) ,
\\
  \vec{j}_q^{Q_k}(\vec{r})
  &=&
  <[\hat{\vec{j}}_q,\hat{Q}_k]>
  \\
  &=&
  \Im\left\{\sum_{n\in q}
  \left(
  \varphi^*_n\nabla\hat{Q}_k\varphi_n^{\mbox{}}
  -
  \hat{Q}_k\varphi^*_n\nabla\varphi_n^{\mbox{}}
  \right)
  \right\} ,
\\
  \vec{\sigma}_q^{Q_k}(\vec{r})
  &=&
   <[\hat{\vec{\sigma}}_q,\hat{Q}_k]>
  \\
  &=&
  \sum_{n\in q}\left(
  \varphi^*_n\hat{\vec{\sigma}}\hat{Q}_k\varphi_n^{\mbox{}}
  -
  \hat{Q}_k\varphi^*_n\hat{\vec{\sigma}}\varphi_n^{\mbox{}}
  \right) .
\end{eqnarray*}

\vspace{2cm} 

{\bf \large FIGURE CAPTIONS}

\vspace{0.5cm}\indent
{\bf Figure 1}: 
Isovector E1 and isoscalar E2 giant resonances in 
$^{40}{\rm Ca}$ and  $^{208}{\rm Pb}$ calculated 
with SkM* forces in full RPA and SRPA at three 
levels of expansion as indicated. The strengths are 
weighted by Lorentz function with the averaging 
parameter $\Gamma =1.0$ MeV.

\vspace{0.5cm}\indent
{\bf Figure 2}:
Isovector E1 resonance in $^{40}{\rm Ca}$ and  
$^{208}{\rm Pb}$
calculated with SkM* forces in full RPA and SRPA with
the complete set $k\!=\!1-4$. The responses
are weighted by Lorentz function with the small
averaging parameter $\Gamma =0.1$ MeV.

\vspace{0.5cm}\indent
{\bf Figure 3}:
Isovector E1 resonance in $^{208}{\rm Pb}$ calculated 
with SkI3 forces in full RPA and SRPA with a set of four 
operators. The responses are weighted by Lorentz function 
with averaging parameter $\Gamma =1.0$ MeV.

\vspace{0.5cm}\indent
{\bf Figure 4}:
Isovector E1 resonance in $^{208}{\rm Pb}$ calculated in
SRPA with the standard set of four operators and using
the SkM* force. Two different phases spaces are compared:
full line = unrestricted space of excitations, dashed 
line = space restricted to $ph$ energies $<30\,{\rm MeV}$ 
as in previous figures.
The responses are weighted by Lorentz function with
averaging parameter $\Gamma =1.0$ MeV.

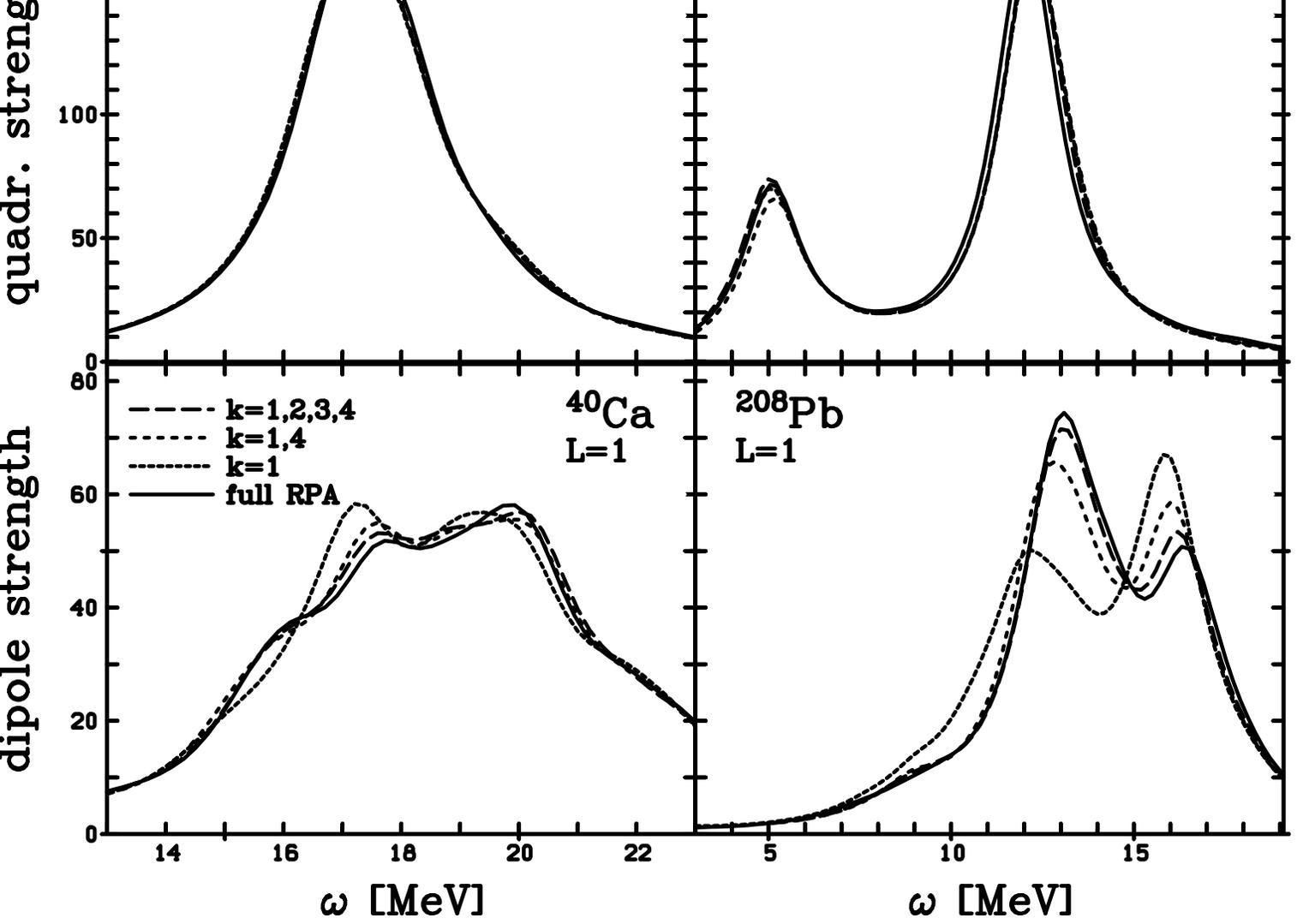
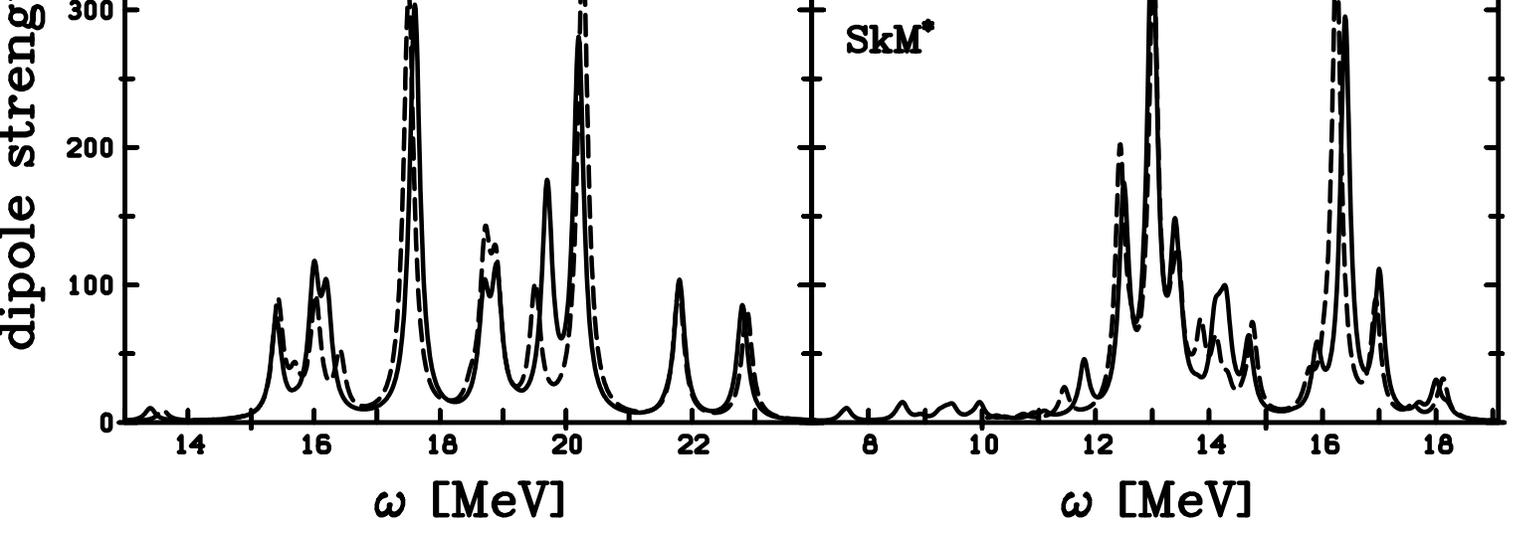
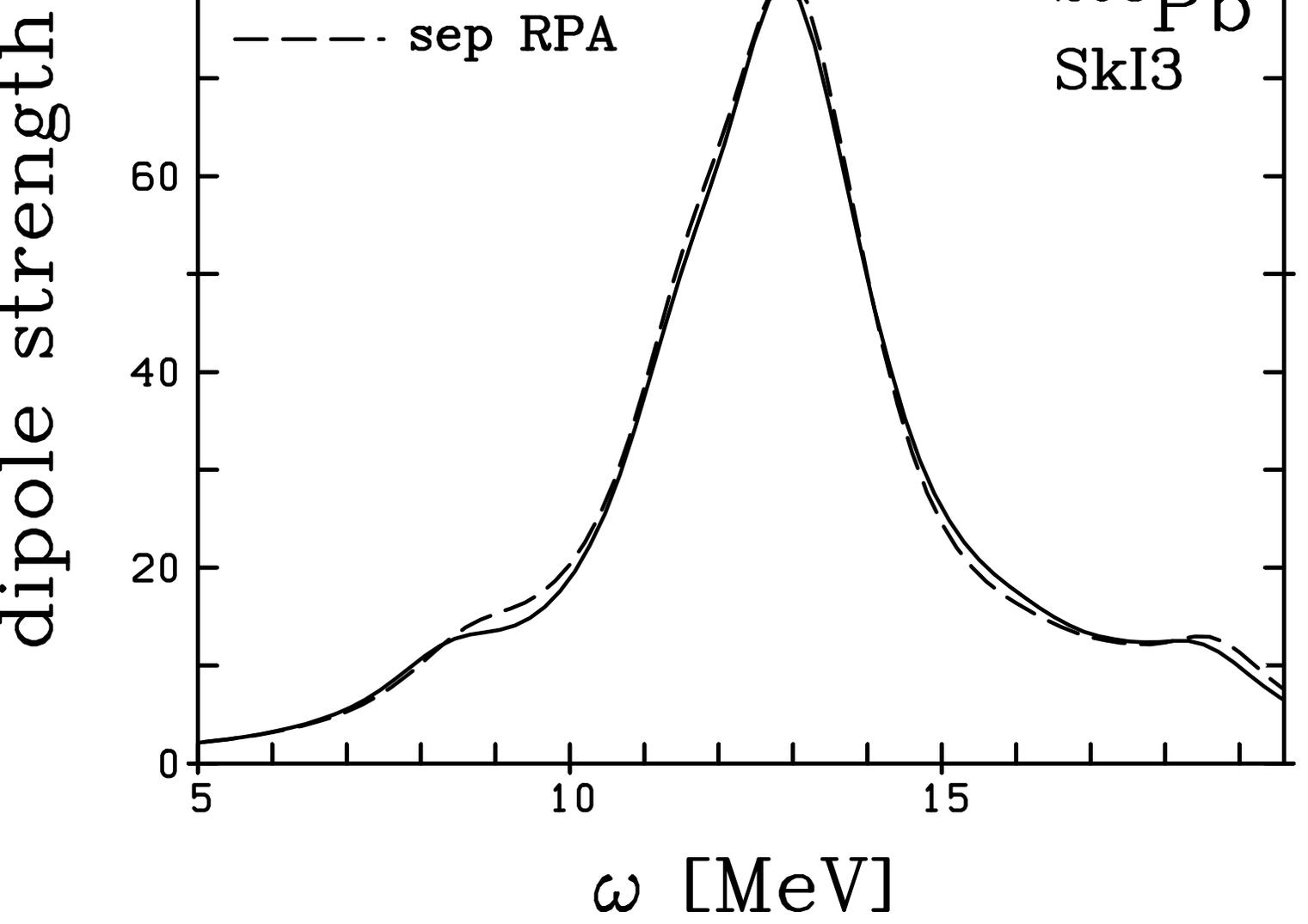
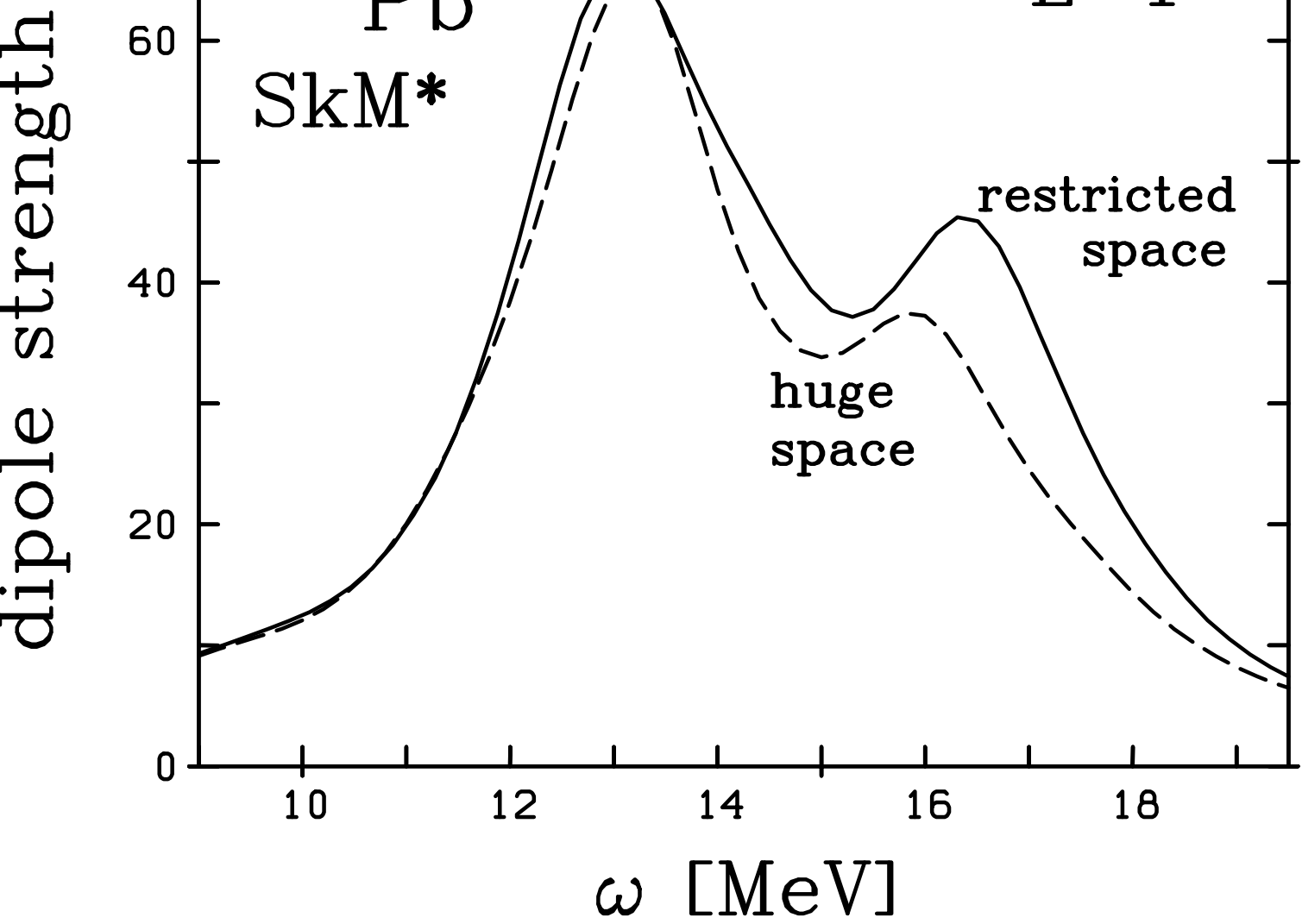
\end{document}